\documentclass[11pt,ams]{article}%
\usepackage{geometry}
\usepackage{amsmath}
\usepackage{amsfonts}
\usepackage{amssymb}
\usepackage{graphicx}%
\providecommand{\U}[1]{\protect\rule{.1in}{.1in}}

\newcommand{\omu}{\overline{\mu}}
\newcommand{\lms}{\Lambda_{\overline{\mbox{\tiny{MS}}}}}

\newcommand{\p}{\partial}
\newcommand{\MSbar}{\overline{\mbox{MS}}}

\begin{document}

\date{}
\title{\textbf{Semiclassical analysis of the phases of $4d$ $SU(2)$ Higgs gauge systems with cutoff at the Gribov horizon}}
\author{\textbf{M.~A.~L.~Capri}$^{a}$\thanks{caprimarcio@gmail.com}\,\,,
\textbf{D.~Dudal}$^{b}$\thanks{david.dudal@ugent.be}\,\,,
\textbf{A.~J.~G\'{o}mez}$^{a}$\thanks{ajgomez@uerj.br}\,\,,
\textbf{M.~S.~Guimaraes}$^{a}$\thanks{msguimaraes@uerj.br}\,\,,
\textbf{I.~F.~Justo }$^{a}$\thanks{igorfjusto@gmail.com}\,\,,\\
\textbf{S.~P.~Sorella}$^{a}$\thanks{sorella@uerj.br}\,\,,
\textbf{D.~Vercauteren}$^{a}$\thanks{vercauteren.uerj@gmail.com}\ \thanks{Work supported by
FAPERJ, Funda{\c{c}}{\~{a}}o de Amparo {\`{a}} Pesquisa do Estado do Rio de
Janeiro, under the program \textit{Cientista do Nosso Estado}, E-26/101.578/2010.}\,\,\\[2mm]
{\small \textnormal{$^{a}$  \it Departamento de F\'{\i }sica Te\'{o}rica, Instituto de F\'{\i }sica, UERJ - Universidade do Estado do Rio de Janeiro,}}
 \\ \small \textnormal{\phantom{$^{a}$} \it Rua S\~{a}o Francisco Xavier 524, 20550-013 Maracan\~{a}, Rio de Janeiro, Brasil}\\
	 \small \textnormal{$^{b}$ \it Ghent University, Department of Physics and Astronomy, Krijgslaan 281-S9, 9000 Gent, Belgium}\normalsize}
\maketitle

\begin{abstract}
We present an analytical study of continuum $4d$ $SU(2)$ gauge Higgs models with a single Higgs field with fixed length in either the fundamental or adjoint representation. We aim at analytically probing the renowned predictions of Fradkin \& Shenker on the phase diagram in terms of confinement versus Higgs behaviour, obtained for the lattice version of the model.  We work in the Landau version of the 't Hooft $R_\xi$ gauges in which case we can access potential nonperturbative physics related to the existence of the Gribov copies. In the fundamental case, we clearly show that in the perturbative regime of small gauge coupling constant $g$ and large Higgs vacuum expectation value $\nu$, there is a Higgs phase with Yukawa gauge boson propagators without Gribov effects. For a small value of the Higgs {\it vev} $\nu$ and/or large $g$, we enter a region with Gribov type propagators that have no physical particle interpretation: the gauge bosons are as such confined. The transition between both behaviours is found to be continuous.  In the adjoint case, we find evidence of a more drastic transition between the different behaviours for the propagator of the off-diagonal gauge bosons, whereas the ``photon'', {\it i.e.}~the diagonal component of the gauge field, displays a propagator of the Gribov type. In the limit of infinite Higgs condensate, we show that a massless photon is recovered. We compare our findings with those of  Fradkin \& Shenker  as well as with more recent  numerical lattice simulations of the fundamental Higgs model. We also carefully discuss in which region of the parameter space $(\nu,g)$ our approximations are trustworthy.

\end{abstract}

\section{Introduction}

The understanding of the transition between the confinement and the Higgs phase of asymptotically free nonabelian gauge theories in presence of Higgs fields is a relevant and yet not fully unraveled topic. Needless to say, such theories are the building blocks of the Standard Model.

Numerical studies of the lattice version of these models \cite{Fradkin:1978dv,Lang:1981qg,Langguth:1985dr,Azcoiti:1987ua,Caudy:2007sf,Bonati:2009pf,Maas:2010nc,Maas:2012zf,Greensite:2011zz} have already revealed a rich structure of the corresponding phase diagram, as well as mean-field approaches to the underlying lattice models as in \cite{Horowitz:1983sr,Damgaard:1985nb,Baier:1986sa}.  What emerges from the lattice works is that the transition between the confinement and the  Higgs phase strongly depends on the representation of the Higgs field.

In the case of the fundamental representation, it turns out that these two phases can be continuously connected, {\it i.e.} they are not completely separated by a transition line. For the benefit of the reader, it is worth spending a few words on this statement. We follow here the seminal work by  \cite{Fradkin:1978dv}, where the lattice version of the model has been considered for a Higgs field with fixed length, {\it i.e.} by freezing its radial part. This amounts to keep the Higgs quartic self-coupling very large, not to say infinite, so that the Higgs field is frozen to its {\it vev}. The resulting theory has two basic parameters, the gauge coupling $g$ and  the {\it vev} $\nu$ of the Higgs field. In the plane $(\nu,g)$, the Higgs phase corresponds to the region of weak coupling, {\it i.e.} small $g$ and sufficiently large $\nu$. In this region, the Wilson loop exhibits a perimeter law and the force between two static sources is short ranged. Instead, the confining phase corresponds to the strong coupling region in the $(\nu,g)$ plane, characterized by large values of $g$ and sufficiently small $\nu$. Here, the Wilson loop gives rise to a linear potential over some finite distance region, followed by string breaking via scalar particle production. It turns out that these two regions are smoothly  connected. More precisely, from e.g.~Fig.1 of \cite{Bonati:2009pf}, one learns that the confining and Higgs phases are separated by a first order line transition which, however, does not extend to the whole phase diagram. Instead, it displays an endpoint. This implies that the Higgs phase can be connected to the confining phase in a continuous way, {\it i.e.} without crossing the transition line. Even more striking, one observes a region in the $(\nu,g)$ plane, called analyticity region \cite{Fradkin:1978dv,Caudy:2007sf,Bonati:2009pf}, in which the Higgs and confining phases are connected by paths along which the expectation value of any local correlation function varies analytically, implying the absence of any discontinuity  in the thermodynamical quantities.  According to  \cite{Fradkin:1978dv}, the spectrum of the theory evolves continuously from one regime to the other. Said otherwise, the physical states in both regimes are generated by suitable gauge invariant operators which, in the Higgs phase, give rise to massive bosons while, in the confining phase, to a meson like state, {\it i.e.} to a bound state of confined excitations. We shall stick to the use of the word ``phase'' here, although it is clear there cannot exist a local order parameter\footnote{Nor can the Polyakov loop $\mathcal{P}$ serve as an order parameter since the presence of a fundamental Higgs field breaks the center symmetry in a hard way, giving anyhow a nontrivial expectation value to $\mathcal{P}$.} discriminating between the Higgs or confinement behaviour.

In the adjoint Higgs case, the breaking or not of center symmetry and associated Polyakov loop expectation value can in principle be used to discriminate between phases, leading to (an) expected phase transition(s) between confining and Higgs behaviour, as already put forward in \cite{Fradkin:1978dv} and tested in e.g.~\cite{Lang:1981qg}. Moreover, as speculated in \cite{Fradkin:1978dv}, the appearance of an in some cases expected Coulomb phase with long-range correlations was not reported. Over the years, most attention has however been paid to the fundamental Higgs, due to its more direct physical relevance for the electroweak physics.

The attentive reader will have noticed that most of the cited works either concern lattice simulations of the gauge Higgs systems, or analytical semiclassical mean-field analysis of the lattice model. There is a good reason for this as the probed physics is usually of a nonperturbative nature. Some preliminary functional Dyson-Schwinger approaches to gauge field theories supplemented with scalar fields have appeared nonetheless \cite{Fister:2010yw,Macher:2011ys}, however these do not really touch upon the supposed phase diagram and the analyticity results of Fradkin \& Shenker. Direct computation of phase diagram related quantities as Polyakov loops or even the vacuum (free) energy are noteworthy hard to access at the functional level.  Let us by the way notice here that the proof of the Fradkin-Shenker analyticity result, inspired by \cite{Osterwalder:1977pc}, heavily relies on the lattice formulation of the problem.  We are unaware of a strict continuum version of the theorem, although one might expect it to hold as well. It is in any case a strong property of the theory if it would be analytical in a certain region of the phase diagram. In some cases, the Fradkin \& Shenker result has been related to the Lee-Yang theorem that also handles analyticity properties of statistical systems \cite{leeyang,Nussinov:2004ns}. One can imagine a lot of subtleties making any concrete analytical study of the Higgs phase diagram troublesome. To name only one, renormalization will require to consider the gauge coupling constant at a particular renormalization scale and the issue of strong versus weak coupling is directly set by the size of this scale. We will come back to this issue later.

The aim of this work is that of investigating the transition between the Higgs and the confinement phase within a continuum quantum field theory. The set up which we shall follow is that of taking into account the nonperturbative effect of the existence of the Gribov copies  \cite{Gribov:1977wm} which are unavoidably present in the gauge fixing quantization procedure\footnote{See \cite{Sobreiro:2005ec,Vandersickel:2012tz}  for a pedagogical introduction to the Gribov problem.}. In particular, as already shown in the case of three-dimensional gauge theories \cite{Capri:2012cr}, this framework enables us to encode nonperturbative information about the phases of the theory in the two-point correlation function of the gauge field in momentum space. The transition between one phase to another is detected by the corresponding change in the pole structure of the gluon propagator. More precisely, a gluon propagateor of the Gribov type, {\it i.e.} displaying complex conjugate poles, has no particle interpretation, being well suited to describe the confining phase. On the other hand, a propagator of the Yukawa type, {\it i.e.} displaying a real pole in momentum space, means that the theory is located in the Higgs phase.

By quantizing the theory in the Landau gauge and by performing the restriction to the so called Gribov region $\Omega$   \cite{Gribov:1977wm,Sobreiro:2005ec,Vandersickel:2012tz} in order to take into account the effect of the Gribov copies, we shall be able to discuss the changes in the gluon propagator when the parameters $(\nu,g)$ are varied from the Higgs weak coupling region to the confining strong coupling region. As we shall see in details, in the case of the fundamental representation we shall be able to detect a region in the $(\nu,g)$ plane, characterized by the line\footnote{The quantity $\omu$ stands for the energy scale which shows up in the renormalization of ultraviolet divergent quantities. In the present case, dimensional regularization in the $\MSbar$ scheme  is employed.}  $a = \frac{1}{2}$, with
\begin{equation}
a = \frac{g^2 \nu^2}{4 \omu^2 e^{\left( 1 -\frac{32\pi^2}{3g^2} \right)}}  \;. \label{aint}
\end{equation}
The Higgs region will correspond to values of $(\nu,g)$  for which $a> \frac{1}{2}$, while the confining phase will be located in the region $a<\frac{1}{2}$. Moreover,  the gluon propagator evolves in a smooth continuous way from the confining to the Higgs region, namely we observe a continuous evolution of the poles of the propagator from a Gribov type to the Yukawa behavior. In addition, we shall be able to evaluate the vacuum energy in  Gribov's semiclassical approximation and check explicitly that it is a continuous function, together with its first and second derivative,  of the parameter $a$ across the transition line $a=\frac{1}{2}$. The third derivative of the  vacuum energy displays a discontinuity across $a=\frac{1}{2}$. However, we shall be able to check that the line $a=\frac{1}{2}$ is not located in the Fradkin-Shenker analyticity region, something which we cannot access within the present approximation as $a=\frac{1}{2}$ corresponds to a region wherein we cannot really trust the made approximations. Though, we are able to confirm within a continuum field theory that the transition between the Higgs and the confining phase occurs in a smooth way, something which is far from being trivial. Another relevant result of our analysis is that, in the Higgs phase, there is no need to implement the restriction  to the Gribov region $\Omega$. Said otherwise, in the Higgs phase, the values of the parameters $(\nu,g)$ are such that the theory automatically lies within the Gribov region $\Omega$, so that the Gribov horizon is never crossed. This is a quite relevant observation which has an important physical consequence.  It ensures that, in the weak coupling region, the standard Higgs mechanism takes place without being affected by the restriction to the Gribov region. We have thus a truly nonconfining theory whose asymptotic states are the massive gauge bosons.

When the Higgs field is in the adjoint representation, our results indicate that things change drastically, as also expected from  lattice investigations. Here, the phase structure looks much more intricate as well as the evolution of the two-point correlation function of the gauge field. In addition of the confining phase, in which the gluon propagator is of the Gribov type, we shall find what can be called a $U(1)$ confining phase, in which the third component  $A^3_\mu$ of the gauge field displays a Gribov type correlation function, while the remaining off-diagonal components, $A^\alpha_\mu$, $\alpha=1,2$,  are of the Yukawa type. It is worth noticing here that this phase has been reported in lattice investigations of the Georgi-Glashow model \cite{Nadkarni:1989na,Hart:1996ac}, {\it i.e.} of three dimensional gauge theories with a Higgs field in the adjoint. Unfortunately, till now, we are unaware of lattice studies of the gluon propagator in four dimensions with Higgs field in the adjoint\footnote{Adjoint lattice gauge-Higgs systems have been studied in e.g.~\cite{Lang:1981qg,Greensite:2004ke} but not directly from the propagator viewpoint.}. So far, only the case of the fundamental representation has been addressed \cite{Maas:2010nc,Maas:2012zf}. From that point of view, we hope that our results will stimulate further studies of the gluon propagator on the lattice in order to confirm the existence of the $U(1)$ confining phase in four dimensions, to the extent that the diagonal gauge boson propagator is not of the massless type for finite value of the Higgs condensate. A lattice study of the adjoint Higgs phase diagram was presented in \cite{Brower:1982yn}, giving evidence of a massless Coulomb phase in the limit of infinite Higgs condensate $\nu$. More precisely, the theory was shown to reduce to a compact $U(1)$ model with its confinement-deconfinement transition. We do however not expect such a transition in the continuum version of QED. We shall see that in the limit $\nu\to\infty$ we can however recover also a massless photon. This is not as trivial a result as it might appear since it involves a delicate cancelation between diverging Higgs condensate and vanishing Gribov parameter.

The paper is organized as follows. In Sect.~2 we discuss the restriction to the Gribov region in the case of the fundamental representation. The behavior of the gluon propagator and of the vacuum energy are discussed within Gribov's approximation. In Sect.~3 we address the more intricate case of the Higgs field in the adjoint representation. Sect.~4 collects our conclusion.

\section{Restriction to the Gribov region $\Omega$ with a fundamental Higgs field}
Let us consider first the case of $SU(2)$ Yang-Mills theories interacting with Higgs fields in the fundamental representation. This will be the most interesting case for future reference as well, when the physical case of $SU(2)\times U(1)$ will be analyzed, {\it i.e.}~the electroweak theory.

Working in Euclidean space and adopting the Landau gauge, $\partial_\mu A^a_\mu=0$, the action of the current model is specified by the following expression
\begin{equation}
S=\int d^{4}x\left(\frac{1}{4}F_{\mu \nu }^{a}F_{\mu \nu }^{a}+
(D_{\mu }^{ij}\Phi ^{j})^{\dagger}( D_{\mu }^{ik}\Phi ^{k})+\frac{\lambda }{2}\left(
\Phi ^{\dagger}\Phi-\nu ^{2}\right) ^{2}+b^{a}\partial _{\mu }A_{\mu
}^{a}+\bar{c}^{a}\partial _{\mu }D_{\mu }^{ab}c^{b}\right)  \;, \label{Sf}
\end{equation}
where the covariant derivative is defined by
\begin{equation}
D_{\mu }^{ij}\Phi^{j} =\partial _{\mu }\Phi^{i} -ig \frac{(\tau^a)^{ij}}{2}A_{\mu }^{a}\Phi^{j} \;.
\end{equation}
The field $b^a$ stands for the Lagrange multiplier implementing the Landau gauge, $\partial_\mu A^a_\mu=0$, while $({\bar c}^a, c^a)$ are the Faddeev-Popov ghosts. The indices $i,j=1,2$ refer to the fundamental representation, and $\tau^a, a=1,2,3$, are the Pauli matrices.  The vacuum configuration which minimizes the energy is achieved by a constant scalar field parameterized as
\begin{equation}
\langle \Phi \rangle  = \left( \begin{array}{ccc}
                                          0  \\
                                          \nu
                                          \end{array} \right)  \;,  \label{vevf}
\end{equation}
It will be understood that we work in the limit $\lambda\to\infty$ for simplicity, {\it  i.e.}~we have a Higgs field frozen at its vacuum expectation value, as in \cite{Fradkin:1978dv}.

All components of the gauge field acquire the same mass $m^2= \frac{g^2\nu^2}{2}$. In fact, for the quadratic part of the action we have now
\begin{equation}
S_{quad}=\int d^{4}x\left( \frac{1}{4} { \left(  \partial_\mu A^a_\nu -\partial_\nu A^a_\mu  \right)} ^2 + b^a \partial_\mu A^a_\mu
+ \frac{g^{2}\nu^{2}}{4}A_{\mu }^{a}A_{\mu }^{a}  \right)  \;. \label{quadf}
\end{equation}
As already mentioned in the Introduction, the aim of the present work is that of analyzing the possible nonperturbative dynamics of the  model by taking into account the Gribov copies. In the Landau gauge\footnote{ It is perhaps worthwhile pointing out here that the Landau gauge is also a special case of the 't Hooft $R_\xi$ gauges, which have proven their usefulness as being renormalizable and offering a way to get rid of the unwanted propagator mixing between (massive) gauge bosons and associated Goldstone modes, $\sim A_\mu \p_\mu \phi$. The latter terms indeed vanish upon using the gauge field transversality. The upshot of specifically using the Landau gauge is that it allows to take into account potential nonperturbative effects related to the gauge copy ambiguity.}, this issue can be faced by restricting the domain of integration in the path integral to the so called Gribov region $\Omega$ \cite{Gribov:1977wm,Sobreiro:2005ec,Vandersickel:2012tz}, defined as the set of all transverse gauge configurations for which the Faddeev-Popov operator is strictly positive, namely
\begin{equation}
\Omega = \;\; \{ A^a_\mu\;, \; \partial_\mu A^a_\mu=0 \;, \; -\partial_\mu D^{ab}_\mu > 0\; \}  \;.  \label{gr}
\end{equation}
The region $\Omega$ is known to be convex and bounded in all directions in field space. The boundary of $\Omega$, where the first vanishing eigenvalue of the Faddeev-Popov operator appears, is called the first Gribov horizon. A way to implement the restriction to the region $\Omega$ has been worked out by Gribov in his original work. It amounts to impose the no-pole condition \cite{Gribov:1977wm,Sobreiro:2005ec,Vandersickel:2012tz} for the connected two-point ghost function $\mathcal{G}^{ab}(k;A) = \langle k | \left(  -\partial D^{ab}(A) \right)^{-1} |k\rangle $, which is nothing but the inverse of the Faddeev-Popov operator $-\partial D^{ab}(A)$. One requires that  $\mathcal{G}^{ab}(k;A)$ has no poles at finite nonvanishing values of $k^2$, so that it stays always positive. In that way one ensures that the Gribov horizon is not crossed, {\it i.e.} one remains inside $\Omega$. The only allowed pole is at $k^2=0$, which has the meaning of approaching the boundary of the region $\Omega$.

Following Gribov's procedure \cite{Gribov:1977wm,Sobreiro:2005ec,Vandersickel:2012tz}, for the connected two-point ghost function $\mathcal{G}^{ab}(k;A)$ at first order in the gauge fields,  one finds
\begin{equation}
\mathcal{G}^{ab}(k;A)=\frac{1}{k^{2}}\left( \delta ^{ab}-g^{2}\frac{k_{\mu
}k_{\nu }}{k^{2}}\int \frac{d^{4}q}{(2\pi )^{4}}\varepsilon
^{amc}\varepsilon ^{cnb}\frac{1}{(k-q)^{2}}\left( A_{\mu }^{m}(q)A_{\nu
}^{n}(-q)\right) \right)  \label{Ghost} \;,
\end{equation}%
where use has been made of the transversality condition $q_{\mu}A_{\mu}(q)=0$. Taking into account that all masses of the gauge field are degenerate in color space, eq.\eqref{quadf}, we introduce the ghost form factor $\sigma(k;A)$ as
\begin{eqnarray}
\mathcal{G}(k;A) &=&   \frac{\delta^{ab} {\cal G}^{ab}}{3} =         \frac{1}{k^{2}}\left( 1+\frac{k_{\mu }k_{\nu }%
}{k^{2}}\frac{2g^2}{3} \int \frac{d^{4}q}{(2\pi )^{4}}\frac{A_{\mu
}^{a }(q)A_{\nu }^{a }(-q)}{(q-k)^{2}}
\right)  \nonumber \\
&\equiv &\frac{1}{k^{2}}\left( 1+\sigma(k;A)\right)       \approx \frac{1}{k^{2}}\left( \frac{1}{1-\sigma(k;A)}\right) \label{sf} \;.  \end{eqnarray}%
The quantity $\sigma(k;A)$ turns out to be a decreasing function of the momentum $k$ \cite{Gribov:1977wm,Sobreiro:2005ec,Vandersickel:2012tz}. Thus, the no-pole condition for the ghost function $\mathcal{G}(k,A)$ is implemented by imposing that \cite{Gribov:1977wm,Sobreiro:2005ec,Vandersickel:2012tz}
\begin{equation}
\sigma(0;A) \leq 1\;,   \label{npf}
\end{equation}%
where $\sigma(0;A)$ is given by
\begin{equation}
\sigma(0;A) =\frac{g^{2}}{6}\int \frac{d^{4}q}{(2\pi )^{4}}\frac{
A_{\mu }^{a}(q)A_{\mu }^{a}(-q) }{q^{2}}  \;. \label{sf1}
\end{equation}
This expression is obtained by taking the limit $k \rightarrow 0$ of eq.\eqref{sf}, and by making use of the property
\begin{eqnarray}
A_{\mu }^{a}(q)A_{\nu }^{a}(-q) &=&\left( \delta _{\mu \nu }-\frac{q_{\mu }q_{\nu }%
}{q^{2}}\right) \omega (A)(q)   \nonumber \\
&\Rightarrow &\omega (A)(q)=\frac{1}{3}A_{\lambda }^{a}(q)A_{\lambda }^{a}(-q)  \label{p1}
\end{eqnarray}
which follows from the transversality of the gauge field, $q_\mu A^a_\mu(q)=0$. Also, it is useful to remind that, for an arbitrary function $\mathcal{F}(p^2)$, we have
\begin{equation}
\int \frac{d^{4}p}{(2\pi )^{4}}\left( \delta _{\mu \nu }-\frac{%
p_{\mu }p_{\nu }}{p^{2}}\right) \mathcal{F}(p^2)=\mathcal{A}\;\delta _{\mu \nu }  \label{p2}
\end{equation}%
where, upon contracting both sides of eq.\eqref{p2} with $\delta_{\mu\nu}$,
\begin{equation}
\mathcal{A}=\frac{3}{4}\int \frac{d^{4}p}{(2\pi )^{4}}\mathcal{F}(p^2).   \label{p3}
\end{equation}%

\subsection{Gribov's  gap equations}
In order to ensure the restriction to the Gribov region $\Omega$ in the functional integral, we encode the information of the no-pole conditions into a step function \cite{Gribov:1977wm,Sobreiro:2005ec,Vandersickel:2012tz}:
\begin{equation}
Z=\int {[dA]\delta (\partial A)\; \det (-\partial D^{ab})\; \theta (1-\sigma(0;A))\; e^{-S_{YM}}.} \label{im}
\end{equation}
Though, as our interest for now lies only in the study of the gauge boson propagators, we shall consider here the quadratic approximation for the partition function, namely  \begin{equation}
Z_{quad} =\int \frac{d\vartheta }{2\pi i\vartheta } %
[dA]\; e^{\vartheta (1-\sigma (0,A))}
\; e^{-\frac{1}{4}\int d^{4}x(\partial _{\mu }A_{\nu }^{a}-\partial
_{\nu }A_{\mu }^{a})^{2}-\frac{1}{2\xi }\int {d^{4}x(\partial _{\mu }A_{\mu
}^{a})^{2}-}\frac{{g^{2}\nu ^{2}}}{4}\int {d^{4}xA_{\mu }^{a }A_{\mu}^{a }}} \;, \label{zf}
\end{equation}
where use has been made of the integral representation
\begin{equation}
\theta(x) = \int_{-i \infty +\epsilon}^{i\infty +\epsilon} \frac{d\vartheta}{2\pi i \vartheta} \; e^{\vartheta x}  \;. \label{stepf}
\end{equation}
The extension of this work beyond the semiclassical approximation can be worked out using the tools of \cite{Vandersickel:2012tz}.  We shall also not dwell upon renormalization details here, we expect that the proof of the pure gauge case as in \cite{Vandersickel:2012tz,Dudal:2010fq} can be suitably adapted.

After simple algebraic manipulations, one gets for \eqref{zf}
\begin{equation}
Z_{quad}=\int \frac{d\vartheta  e^{\vartheta  }}{2\pi i\vartheta } [dA] \; e^{-\frac{1}{2} \int \frac{%
d^{4}q}{(2\pi )^{4}}A_{\mu }^{a }(q)\mathcal{P}_{\mu \nu }^{ab }A_{\nu }^{b }(-q)},  \label{Zq1f}
\end{equation}
with
\begin{equation}
\mathcal{P}_{\mu \nu }^{ab } =\delta ^{ab }\left(
\delta _{\mu \nu }\left( q^{2}+\frac{\nu^{2}g^{2}}{2}\right) +\left( \frac{1}{\xi }%
-1\right) q_{\mu }q_{\nu }+\frac{\vartheta}{3} \frac{g^{2}}{q^2} \delta _{\mu \nu }\right).  \label{Pf}
\end{equation}
The parameter $\xi$ is understood to be zero at the end in order to recover the
Landau gauge. Evaluating  the inverse of the
expression above and taking the limit $\xi\rightarrow 0$, for the gluon propagator  one gets
\begin{equation}
\left\langle A_{\mu }^{a}(q)A_{\nu }^{b}(-q)\right\rangle
=\delta^{ab}\frac{ q^2}{q^{4} + \frac{g^{2}\nu^{2}}{2} q^2
+ \frac{g^2}{3} \vartheta }\left( \delta _{\mu
\nu }-\frac{q_{\mu }q_{\nu }}{q^{2}}\right)  \label{propf} \;.
\end{equation}
It remains to find the gap equation for the Gribov parameter $\vartheta$, enabling us to express it in terms of the parameters of the starting model, {\it i.e.} the gauge coupling constant $g$ and the {\it vev} of the Higgs field $\nu$. In order to accomplish this task we follow \cite{Gribov:1977wm,Sobreiro:2005ec,Vandersickel:2012tz} and evaluate the partition function $Z_{quad}$ in the semiclassical approximation. First, we integrate out the gauge fields, obtaining
\begin{equation}
Z_{quad}=\int {\frac{d\vartheta}{2\pi i} }
e^{({\vartheta} -\ln\vartheta)} \left(\det\mathcal{P}^{ab}_{\mu\nu}\right)^{-\frac{1}{2}} \;. \label{Zq2f}
\end{equation}
Making use of
\begin{equation}
\left(\det \mathcal{P}_{\mu\nu}^{ab}\right)^{-\frac{1}{2}}=e^{-\frac{1}{2}%
\ln \det \mathcal{P}_{\mu\nu}^{ab}}=e^{-\frac{1}{2}Tr \ln \mathcal{P}%
_{\mu\nu}^{ab}} \;,
\end{equation}
for the determinant in expression (\ref{Zq2f}) we get
\begin{equation}
\left( \det \mathcal{P}_{\mu \nu }^{ab }\right) ^{-\frac{1}{2}} = \exp \left[ -\frac{9}{2}\int {\frac{d^{4}q}{(2\pi )^{4}}\ln \left(
q^{2}+\frac{g^{2}\nu^{2}}{2}+\frac{g^{2} \vartheta}{3}
\frac{1}{q^{2}}\right) }\right] \;.
\end{equation}
Therefore,
\begin{equation}  \label{Zf}
Z_{quad}=\int \frac{d\vartheta}{2\pi i}\; e^{f(\vartheta)} \;,
\end{equation}
where, in the thermodynamic limit\footnote{We remind here that the term ${\ln\vartheta}$  in expression \eqref{Zq2f} can be neglected in the derivation of the gap equation, eq.\eqref{gapf}, when taking the thermodynamic limit, see \cite{Gribov:1977wm,Sobreiro:2005ec,Vandersickel:2012tz} for details.},
\begin{equation}
f(\vartheta)= \vartheta - \frac{9}{2} \int \frac{d^4k}{(2\pi)^4} \; \ln\left( k^2 + \frac{g^2\nu^2}{2} +\frac{\vartheta}{3} \frac{g^2}{k^2} \right)  \;. \label{ff}
\end{equation}
Expression (\ref{Zf}) can be now evaluated in the saddle point approximation \cite{Gribov:1977wm,Sobreiro:2005ec,Vandersickel:2012tz}, {\it i.e.}
\begin{equation}
Z_{quad}\approx e^{f(\vartheta^*)} \;,  \label{ef}
\end{equation}
where the parameter $\vartheta^*$ is determined by the stationary condition
\begin{equation}
\frac{\partial f}{\partial \vartheta^*} =0 \;,
\end{equation}
which yields the following gap equation
\begin{equation}
\frac{3}{2}g^2 \int \frac{d^4q}{(2\pi)^4} \frac{1}{ q^{4} + \frac{g^{2}\nu^{2}}{2} q^2
+ \frac{g^2}{3}\vartheta^*}  = 1  \;. \label{gapf}
\end{equation}
Notice also that the function $f(\vartheta^*)$ has the meaning of the vacuum energy ${\cal E}_v$ of the system. More precisely
\begin{equation}
{\cal E}_v = - f(\vartheta^*)   \;, \label{ve}
\end{equation}
as it is apparent from the expression of the partition function $Z_{quad}$, eq.\eqref{ef}. To discuss the gap equation \eqref{gapf}, we decompose the denominator according to
\begin{equation}
q^4 +  \frac{g^{2}\nu^{2}}{2} q^2
+ \frac{g^2}{3} \vartheta   = (q^2+m^2_+) (q^2+m^2_-) \;,  \label{dec1}
\end{equation}
with
\begin{equation}
m^2_+ = \frac{1}{2} \left(\frac{g^2 \nu^2}{2} + \sqrt{\frac{g^4\nu^4}{4}  -\frac{4g^2}{3} \vartheta^*} \;  \right) \;,  \qquad    m^2_- = \frac{1}{2} \left(\frac{g^2 \nu^2}{2} -\sqrt{\frac{g^4\nu^4}{4}  -\frac{4g^2}{3} \vartheta^*} \;  \right)  \;.
\label{roots1}
\end{equation}
Making use of the $\MSbar$ renormalization scheme in $d=4-\varepsilon$ and of the standard integral
\begin{equation}
\int \frac{d^dp}{(2\pi)^d} \frac{1}{p^2+\rho^2} = - \frac{\rho^2}{16\pi^2} \frac{2}{\bar \varepsilon} + \frac{\rho^2}{16\pi^2} \left( \ln\frac{\rho^2}{{\omu}^2} - 1  \right) \;,\label{intd2f}
\end{equation}
for the gap equation \eqref{gapf} we consequently get
\begin{equation}
\left(1 + \frac{m^2_-}{m^2_+ -  m^2_-}\; \ln\left( \frac{m^2_-}{{\omu}^2} \right)  - \frac{m^2_+}{m^2_+ -  m^2_-}\; \ln\left( \frac{m^2_+}{{\omu}^2} \right)  \right) = \frac{32\pi^2}{3g^2}  \;. \label{gapf1}
\end{equation}
In order to analyze this equation we rewrite it in a more suitable way, {\it i.e.}
\begin{equation}
 \frac{m^2_-}{m^2_+ -  m^2_-}\; \ln\left( \frac{m^2_-}{\omu^2} \right)  - \frac{m^2_+}{m^2_+ -  m^2_-}\; \ln\left( \frac{m^2_+}{\omu^2} \right)   = - \frac{m^2_+ - m^2_-}{m^2_+ -  m^2_-} \left( 1- \frac{32\pi^2}{3g^2} \right)  =  \frac{m^2_+ - m^2_-}{m^2_+ -  m^2_-}\; \ln \left(  e^{- \left( 1- \frac{32\pi^2}{3g^2} \right)}  \right)   \;,  \label{gapf2}
\end{equation}
so that
\begin{equation}
 m^2_-\; \ln\left( \frac{m^2_-}{\omu^2  e^{\left( 1- \frac{32\pi^2}{3g^2} \right)} } \right) =  m^2_+\; \ln\left( \frac{m^2_+}{\omu^2  e^{\left( 1- \frac{32\pi^2}{3g^2} \right)} } \right) \;,
  \label{gapf3}
\end{equation}
whose final form can be written as
\begin{equation}
2 \sqrt{1-\zeta}\; \ln(a) = -  \left( 1 +  \sqrt{1-\zeta} \right) \; \ln\left( 1 +  \sqrt{1-\zeta} \right) +  \left( 1 -  \sqrt{1-\zeta} \right) \; \ln\left( 1 - \sqrt{1-\zeta} \right)  \;, \label{gapd1}
\end{equation}
where we have introduced the dimensionless variables
\begin{equation}
a = \frac{g^2 \nu^2}{4 \omu^2 e^{\left( 1 -\frac{32\pi^2}{3g^2} \right)}} \;, \qquad  \qquad \zeta = \frac{16}{3} \frac{\vartheta^*}{g^2\nu^4}  \geq0 \;, \label{vb1}
\end{equation}
with $0 \le \zeta < 1$ in order to have two real, positive, distinct roots $(m^2_+, m^2_-)$, eq.\eqref{roots1}. It is worth to underline that the renormalization scale $\omu$ could be exchanged in favour of the invariant scale  $\lms$, defined at one-loop as
\begin{equation}
\lms^2  =  \omu^2 \;e^{\frac{1}{\beta_0} \frac{1}{ g^2(\omu) } }  \label{Lambda} \;,
\end{equation}
with $\beta_0$ given by \cite{Gross:1973ju,Pickering:2001aq}
\begin{equation}
\beta  = - g^3 \beta_0 + O(g^5) \;, \qquad  \beta_0 = \frac{1}{16\pi^2} \left( \frac{11}{3} N - \frac{1}{6} T  \right)  \;,
\end{equation}
where $T$ is the Casimir of the representation of the Higgs field equaling $T=\frac{1}{2}$, resp.~$T=2$, for the fundamental, resp.~adjoint, representation of $SU(2)$.

For $\zeta >1$, the roots $(m^2_+, m^2_-)$ become complex conjugate, and the gap equation takes the form
\begin{equation}
2 \sqrt{\zeta -1} \; \ln(a) =   -2 \; \arctan\left({\sqrt{\zeta-1}}\; \right)   - \sqrt{\zeta-1} \; \ln\;\zeta  \;. \label{c1}
\end{equation}
Moreover, it is worth noticing that both expressions \eqref{gapd1},\eqref{c1} involve only one function, {\it i.e.} they can be written as
\begin{equation}
2 \; \ln(a) = g(\zeta)  \;, \label{g}
\end{equation}
where for $g(\zeta)$ we might take
\begin{equation}
g(\zeta) =    \frac{1}{ \sqrt{1-\zeta}} \left(
- \left( 1 +  \sqrt{1-\zeta} \right) \; \ln\left( 1 +  \sqrt{1-\zeta} \right) +  \left( 1 -  \sqrt{1-\zeta} \right) \; \ln\left( 1 - \sqrt{1-\zeta} \right)  \right) \;,
\label{gex}
\end{equation}
which is a real function of the variable $\zeta \ge 0$. Expression \eqref{c1} is easily obtained from \eqref{gapd1} by rewriting it in the region $\zeta>1$.  In particular, it turns out that the function $g(\zeta)\leq -2\ln 2$ for all $\zeta \ge 0$, and strictly decreasing. As consequence, for each value of $a<\frac{1}{2}$, equation \eqref{g} has always a unique solution with $\zeta>0$.  Moreover, it is easy to check that $g(1)=-2$. Therefore,  we can distinguish ultimately three regions, namely
\begin{itemize}
\item[(a)]  when $a>\frac{1}{2}$,  eq.\eqref{g}  has no solution for $\zeta$. As the gap equation \eqref{gapf} has been  obtained by acting with $\frac{\p}{\p \vartheta}$ on the expression of the vacuum energy ${\cal E}_v = - f(\vartheta)$, eq.\eqref{ve}, we are forced to set $\vartheta=0$.  This means that, when $a>\frac{1}{2}$, the dynamics of the system is such that the restriction to the Gribov region cannot be consistently implemented.  As a consequence, the standard Higgs mechanism takes place, yielding three massive gauge fields, according to
\begin{equation}
\left\langle A_{\mu }^{a}(q)A_{\nu }^{b}(-q)\right\rangle
=\delta^{ab}\frac{1}{q^{2} + \frac{g^{2}\nu^{2}}{2}  }\left( \delta _{\mu
\nu }-\frac{q_{\mu }q_{\nu }}{q^{2}}\right)  \label{propfY} \;.
\end{equation}
For sufficiently weak coupling $g^2$, we underline that $a$ will unavoidably be larger than $\frac{1}{2}$.

\item[(b)] when $\frac{1}{e}<a<\frac{1}{2}$, equation  \eqref{g} has a solution for  $0 \le \zeta <1$. In this region, the roots $(m^2_+, m^2_-)$  are real and the gluon propagator decomposes into the sum of two terms of the Yukawa type:
\begin{equation}
\left\langle A_{\mu }^{a }(q)A_{\nu }^{b }(-q)\right\rangle
=\delta ^{\alpha \beta } \left(  \frac{{\cal F}_+}{q^2+m^2_+} -    \frac{{\cal F}_-}{q^2+m^2_-}   \right)
\left( \delta _{\mu
\nu }-\frac{q_{\mu }q_{\nu }}{q^{2}}\right)  \label{ffin} \;,
\end{equation}
where
\begin{equation}
{\cal F}_+ = \frac{m^2_+}{m^2_+-m^2_-}  \;, \qquad {\cal F}_- = \frac{m^2_-}{m^2_+-m^2_-} \;. \label{rf}
\end{equation}
Moreover, due to the relative minus sign in eq.\eqref{ffin} only the component ${\cal F}_+$ represents a physical mode.
\item[(c)]  for $a<\frac{1}{e}$, equation \eqref{g} has a solution for  $\zeta>1$. This scenario will always be realized if $g^2$ gets sufficiently large, i.e.~at strong coupling. In this region the roots  $(m^2_+, m^2_-)$  become complex conjugate and the gauge boson propagator is of the Gribov type, displaying complex poles.  As usual, this can be interpreted as the confining region.
\end{itemize}
In summary, we clearly notice that at sufficiently weak coupling, the standard Higgs mechanism, eq.\eqref{propfY}, will definitely take place, as $a>\frac{1}{2}$, whereas for sufficiently strong coupling, we always end up in a confining phase because then $a<\frac{1}{2}$.

Having obtained these results, it is instructive to go back where we originally started. For a fundamental Higgs, all gauge bosons acquire a mass that screens the propagator in the infrared. This effect, combined with a sufficiently small coupling constant, will lead to a severely suppressed ghost self energy, i.e.~the average of \eqref{sf1} (to be understood after renormalization of course). If the latter quantity will a priori not exceed the value of 1 under certain conditions, the theory is already well inside the Gribov region and there is no need to implement the restriction. Actually, the failure of the Gribov restriction for $a>\frac{1}{2}$ is exactly because it is simply not possible to enforce that $\sigma(0)=1$. Perturbation theory in the Higgs sector is \emph{in se} already consistent with the restriction within the 1st Gribov horizon. Let us verify this explicitly by taking the average of \eqref{sf1} with as tree level input propagator a transverse Yukawa one with mass $m^2=\frac{g^2\nu^2}{2}$, cfr.~eq.~\eqref{quadf}. Using that there are 3 transverse directions\footnote{We have been a bit sloppy in this paper with the use of dimensional regularization. In principle, there are $3-\epsilon$ transverse polarizations in $d=4-\epsilon$ dimensions. Positive powers in $\epsilon$ can (and will) combine with the divergences in $\epsilon^{-1}$ to change the finite terms. However, as already pointed out before, a careful renormalization analysis of the Gribov restriction is possible, see e.g.~\cite{Dudal:2010fq,Vandersickel:2012tz} and this will also reveal that the ``1'' in the Gribov gap equation will receive finite renormalizations, compatible with the finite renormalization in e.g.~$\sigma(0)$, basically absorbable  in the definition $a$. The main results of our current paper thus remain correct and we leave the full renormalization details for later. } in $4d$, we easily get
\begin{eqnarray}
\sigma(0) =\frac{3g^{2}}{2}\int \frac{d^{4}q}{(2\pi )^{4}}\frac{1}{q^2(q^2+\frac{g^2\nu^2}{2})}=-\frac{3}{\nu^2}\int\frac{d^4q}{(2\pi^4)}\frac{1}{q^2+\frac{g^2\nu^2}{2}}=-\frac{3g^2}{32\pi^2}\left(\ln\frac{g^2\nu^2}{2\omu^2}-1\right)
\end{eqnarray}
Introducing $a$ as in \eqref{vb1}, we may reexpress the latter result as
\begin{eqnarray}
\sigma(0) =1-\frac{3g^2}{32\pi^2}\ln(2a)
\end{eqnarray}
For $a>\frac{1}{2}$, the logarithm is positive and it is then evident that $\sigma(0)$ will not cross $1$, indicating that the theory already is well within the first Gribov horizon.

Another interesting remark is at place concerning the transition in terms of a varying value of $a$. If $a$ crosses $\frac{1}{e}$, the imaginary part of the complex conjugate roots becomes smoothly zero, leaving us with 2 coinciding real roots, which then split when $a$ grows. At $a=\frac{1}{2}$, one of the roots and its accompanying residue vanishes, to leave us with a single massive gauge boson. We thus observe all these transitions are continuous, something which is in qualitative correspondence with the theoretical lattice predictions of the classic work \cite{Fradkin:1978dv} for a fundamental Higgs field that is ``frozen'' ($\lambda\to\infty$). Concerning the somewhat strange intermediate phase, {\it i.e.}  the one with a Yukawa propagator with a negative residue, eq.\eqref{ffin}, we can investigate in future work in more detail the asymptotic spectrum based on the BRST tools developed in \cite{Dudal:2012sb} when the local action formulation of the Gribov restriction is implemented.

\subsection{The vacuum energy, phase transition and how trustworthy are the results?}
Let us look at the vacuum energy ${\cal E}_v$ of the system, which  can be easily read off from expression \eqref{Zf}, namely
\begin{equation}
{\cal E}_v = - \vartheta^* + \frac{9}{2} \int \frac{d^4k}{(2\pi)^4} \; \ln\left( k^2 + \frac{g^2\nu^2}{2} +\frac{\vartheta^*}{3} \frac{g^2}{k^2} \right)  \;, \label{ev}
\end{equation}
where $\vartheta^*$ is given by the gap equation \eqref{gapf}. Making use of
\begin{equation}
\int \frac{d^dp}{(2\pi)^d} \; \ln(p^2 + m^2)  = - \frac{m^4}{32\pi^2} \left( \frac{2}{\bar \varepsilon}  -  \ln{\frac{m^2}{{\bar \mu}^2}} + \frac{3}{2} \right) \;,  \label{intdl}
\end{equation}it is very easy to write down the vacuum energy:
\begin{itemize}
\item  for $a<\frac{1}{2}$, we have
\begin{eqnarray}
\frac{8}{9 g^4\nu^4}\; {\cal E}_v & = &  \frac{1}{32\pi^2} \left( 1 - \frac{32\pi^2}{3g^2} \right) - \frac{1}{2} \frac{\zeta}{32\pi^2} + \frac{1}{4}\frac{1}{32\pi^2} \left(  (4-2\zeta)\left( \ln( a) -\frac{3}{2} \right) \right)  \\ \nonumber & + &  \frac{1}{4}\frac{1}{32\pi^2} \left(   \left( 1+ \sqrt{1-\zeta} \right)^2 \ln  \left( 1+ \sqrt{1-\zeta} \right)
+ \left( 1- \sqrt{1-\zeta} \right)^2 \ln  \left( 1- \sqrt{1-\zeta} \right)
 \right)     \;, \label{v1}
\end{eqnarray}
where $\zeta$ is obtained through eqs.\eqref{g},\eqref{gex}. Let us also give the expressions of the first two derivatives of ${\cal E}_v(a)$  with respect to $a$. From the gap equation \eqref{g},  we easily get
\begin{equation}
\frac{\partial \zeta}{\partial a} = \frac{2}{a} \frac{1}{g'(\zeta)}  \;. \label{saa}
\end{equation}
Therefore
\begin{eqnarray}
\frac{ \partial }{\partial a} \left[\frac{8}{9 g^4\nu^4}\; {\cal E}_v \right] & = & \frac{1}{64\pi^2} \frac{1}{a} (2-\zeta) \;, \nonumber \\
\frac{ \partial^2 }{\partial a^2} \left[\frac{8}{9 g^4\nu^4}\; {\cal E}_v \right] & =& \frac{1}{64\pi^2} \frac{1}{a^2} \left(\zeta - 2 - \frac{2}{g'({\zeta})} \right)\;. \label{d12}
\end{eqnarray}

\item for $a>\frac{1}{2}$,
\begin{equation}
\frac{8}{9 g^4\nu^4}\; {\cal E}_v  =   \frac{1}{32\pi^2} \left( 1 - \frac{32\pi^2}{3g^2} \right)  + \frac{1}{32\pi^2} \left(  \left( \ln( a) -\frac{3}{2} \right) \right)   +   \frac{1}{32\pi^2} \ln2    \;. \label{v2}
\end{equation}
\end{itemize}
Owing to the fact that $g'(0) =- \infty$, it turns out that vacuum energy ${\cal E}_v(a)$ is a continuous  function of the variable $a$, as well as its first and second derivative. The third derivative develops a jump at $a=\frac{1}{2}$. We might be tempted to interpret this is indicating a third order phase transition at $a=\frac{1}{2}$. The latter value actually corresponds to a line in the $(g^2,\nu)$ plane according to the functional relation \eqref{vb1}. However, we should be cautious to blindly interpret this value. It is important to take a closer look at the validity of our results in the light of the made assumptions. More precisely, we implemented the restriction to the horizon in a first order approximation, which can only be meaningful if the effective coupling constant is sufficiently small, while simultaneously emerging logarithms should be controlled as well. In the absence of propagating matter, the expansion parameter is provided by $y\equiv\frac{g^2N}{16\pi^2}$ as in pure gauge theory. The size of the logarithmic terms in the vacuum energy (that ultimately defines the gap equations) are set by $m_+^2\ln\frac{m_+^2}{\omu^2}$ and $m_-^2\ln\frac{m_-^2}{\omu^2}$. A good choice for the renormalization scale would thus be $\omu^2\sim |m_+^2|$: for (positive) real masses, a fortiori we have $m_-^2<m_+^2$ and the second log will not get excessively large either because $m_-^2$ gets small and the prefactor is thus small, or $m_-^2$ is of the order of $m_+^2$ and the log itself small. For complex conjugate masses, the size of the log is set by the (equal) modulus of $m_\pm^2$ and thus both small by our choice of scale.

Let us now consider the trustworthiness, if any, of the $a=\frac{1}{2}$ phase transition point.  For $a\sim\frac{1}{2}$, we already know that $\zeta\sim 0$, so a perfect choice is $\omu^2\sim m_+^2\sim\frac{g^2\nu^2}{2}$. Doing so, the $a$-equation corresponds to
\begin{equation}\label{aeq}
    \frac{1}{2}\sim e^{-1+\frac{4}{3y}}
\end{equation}
so that $y\sim 4$. Evidently, this number is thus far too big to associate any meaning to the ``phase transition'' at $a=\frac{1}{2}$. Notice that there is no problem for the $a$ small and $a$ large region. If $\nu^2$ is sufficiently large and we set $\omu^2\sim \frac{g^2\nu^2}{2}$ we have a small $y$, leading to a large $a$, i.e.~the weak coupling limit without Gribov parameter and normal Higgs-like physics. The logs are also well-tempered.    For a small $\nu^2$, the choice $\omu^2\sim\sqrt{g^2\theta^*}$ will lead to
\begin{equation}\label{aeq2}
a\sim (\textrm{small number})e^{-1+\frac{4}{3y}}
\end{equation}
so that a small $a$  can now be compatible with a small $y$, leading to a Gribov parameter dominating the Higgs induced mass, the ``small number'' corresponds to $\frac{g^2\nu^2}{\sqrt{g^2\theta^*}}$. Due to the choice of $\omu^2$, the logs are again under control in this case.

Within the current approximation, we are thus forced to conclude that only for sufficiently small or large values of the parameter $a$ we can probe the theory in a controllable fashion. Nevertheless, this is sufficient to ensure the existence of a Higgs-like phase at large Higgs condensate, and a confinement-like region for small Higgs condensate. The intermediate $a$-region is more difficult to interpret due to the occurrence  of large logs and/or effective coupling. Notice that this also might make the emergence of this double Yukawa phase at $a=\frac{1}{e}\approx 0.37$ not well established at this point.

\section{Restriction to the Gribov region $\Omega$ with an adjoint Higgs field}
Let us face now the case in which the  Higgs field $\Phi^a$ transforms according to the adjoint representation  of $SU(2)$ . For the action, we now have
\begin{equation}
S=\int d^{4}x\left(\frac{1}{4}F_{\mu \nu }^{a}F_{\mu \nu }^{a}+\frac{1}{2}%
D_{\mu }^{ab}\Phi ^{b}D_{\mu }^{ac}\Phi ^{c}+\frac{\lambda }{2}\left(
\Phi ^{a}\Phi ^{a}-\nu ^{2}\right) ^{2}+b^{a}\partial _{\mu }A_{\mu
}^{a}+\bar{c}^{a}\partial _{\mu }D_{\mu }^{ab}c^{b}\right)  \;, \label{S}
\end{equation}
where the covariant derivative is defined by
\begin{equation}
\left( D_{\mu }\Phi \right) ^{a}=\partial _{\mu }\Phi ^{a}+g\epsilon
^{abc}A_{\mu }^{b}\Phi ^{c} \;.
\end{equation}
The vacuum configuration which minimizes the energy is achieved by a constant scalar field satisfying
\begin{equation}
\Phi ^{a}\Phi ^{a}=\nu ^{2} \;.
\end{equation}
Setting
\begin{equation}
\left\langle \Phi ^{a}\right\rangle=\nu \delta ^{a3}\;, \label{higgsv}
\end{equation}
for the quadratic part of the action involving the gauge field $A^a_\mu$,  one gets
\begin{equation}
S_{quad}=\int d^{3}x\left( \frac{1}{4} { \left(  \partial_\mu A^a_\nu -\partial_\nu A^a_\mu  \right)} ^2 + b^a \partial_\mu A^a_\mu
+ \frac{g^{2}\nu ^{2}}{2}\left( A_{\mu }^{1}A_{\mu }^{1}+A_{\mu }^{2}A_{\mu
}^{2}\right)  \right)  \;, \label{quad}
\end{equation}
from which one notices that the off-diagonal components $A^{\alpha}_\mu, \alpha=1,2$ acquire a mass $g^2\nu^2$. We again take $\lambda\to\infty$.

In order to implement the restriction to the Gribov region $\Omega$, we start from the expression of the two point ghost function $\mathcal{G}^{ab}(k;A)$ of eq.\eqref{Ghost}. In contrast with the case of the fundamental representation, in order to correctly take into account the presence of the Higgs vacuum, eq.\eqref{higgsv}, we must decompose  $\mathcal{G}^{ab}(k;A)$ into its diagonal and off-diagonal components, according to
\begin{equation}
\mathcal{G}^{ab}(k,A)=\left(
\begin{array}{cc}
\delta^{\alpha \beta}\mathcal{G}_{off}(k;A) & 0 \\
0 & \mathcal{G}_{diag}(k;A)
\end{array}
\right)
\end{equation}
where
\begin{eqnarray}
\mathcal{G}_{off}(k;A) &=&\frac{1}{k^{2}}\left( 1+g^{2}\frac{k_{\mu }k_{\nu }%
}{2k^{2}}\int \frac{d^{4}q}{(2\pi )^{4}}\frac{1}{(q-k)^{2}}\left( A_{\mu
}^{\alpha }(q)A_{\nu }^{\alpha }(-q)+2A_{\mu }^{3}(q)A_{\nu }^{3}(-q)\right)
\right)  \nonumber \\
&\equiv &\frac{1}{k^{2}}\left( 1+\sigma _{off}(k;A)\right)       \approx \frac{1}{k^{2}}\left( \frac{1}{1-\sigma
_{off}(k;A)}\right) \label{Goff} \;,  \\[5mm]
\mathcal{G}_{diag}(k;A) &=&\frac{1}{k^{2}}\left( 1+g^{2}\frac{k_{\mu }k_{\nu
}}{k^{2}}\int \frac{d^{4}q}{(2\pi )^{4}}\frac{1}{(q-k)^{2}}\left( A_{\mu
}^{\alpha }(q)A_{\nu }^{\alpha }(-q)\right) \right)  \nonumber \\
&\equiv &\frac{1}{k^{2}}\left( 1+\sigma _{diag}(k;A)\right)  \approx \frac{1}{k^{2}}\left( \frac{1}{1-\sigma
_{diag}(k;A)}\right) \label{Gdiag} \;.
\end{eqnarray}%
The quantities $\sigma_{off}(k;A), \; \sigma_{diag}(k;A)$ turn out to be  decreasing functions of the momentum $k$ \cite{Gribov:1977wm,Sobreiro:2005ec,Vandersickel:2012tz}. Thus, in the case of the adjoint representation,  the no-pole condition for the ghost function $\mathcal{G}^{ab}(k,A)$ is implemented by demanding that \cite{Gribov:1977wm,Sobreiro:2005ec,Vandersickel:2012tz}
\begin{eqnarray}
\sigma _{off}(0;A) &\leq &1\;, \nonumber \\
\sigma _{diag}(0;A) &\leq &1   \label{np} \;,
\end{eqnarray}%
where $\sigma_{off}(0;A), \; \sigma_{diag}(0;A)$ are given by
\begin{eqnarray}
\sigma _{off}(0;A) &=&\frac{g^{2}}{4}\int \frac{d^{4}q}{(2\pi )^{4}}\frac{%
\left( A_{\mu }^{3}(q)A_{\mu }^{3}(-q)+\frac{1}{2}A_{\mu }^{\alpha
}(q)A_{\mu }^{\alpha }(-q)\right) }{q^{2}}  \;, \nonumber  \\
\sigma _{diag}(0;A) &=&\frac{g^{2}}{4}\int {\frac{d^{4}q}{(2\pi )^{4}}\frac{%
\left( A_{\mu }^{\alpha }(q)A_{\mu }^{\alpha }(-q)\right) }{q^{2}}   \label{sigma} \;, }
\end{eqnarray}
where use has been made of eqs.\eqref{p1},\eqref{p2},\eqref{p3}.

To implement the restriction to the Gribov region $\Omega$ in the functional integral, we proceed as before and encode the no-pole conditions, eqs.\eqref{np}, into step functions \cite{Gribov:1977wm,Sobreiro:2005ec,Vandersickel:2012tz}, obtaining
\begin{eqnarray}
Z_{quad} &=&\int \frac{d\beta }{2\pi i\beta }\frac{d\omega }{2\pi i\omega }%
[dA]\;e^{\beta (1-\sigma _{diag}(0,A))}e^{\omega \left( 1-\sigma
_{off}(0,A)\right) }  \nonumber  \label{Zq} \\
&\times &e^{-\frac{1}{4}\int d^{4}x(\partial _{\mu }A_{\nu }^{a}-\partial
_{\nu }A_{\mu }^{a})^{2}-\frac{1}{2\xi }\int {d^{4}x(\partial _{\mu }A_{\mu
}^{a})^{2}-}\frac{{g^{2}\nu ^{2}}}{2}\int {d^{4}xA_{\mu }^{\alpha }A_{\mu
}^{\alpha }}} \;.
\end{eqnarray}
Notice that, in the case of the adjoint representation, two Gribov parameters $(\beta, \omega)$ are needed in order to implement conditions \eqref{np}, see also the three dimensional case recently discussed in \cite{Capri:2012cr}. After simple algebraic manipulations, we get
\begin{equation}
Z_{quad}=\int \frac{d\beta e^{\beta }}{2\pi i\beta }\frac{d\omega e^{\omega }%
}{2\pi i\omega }[dA^{\alpha }][dA^{3}]\;e^{-\frac{1}{2}\int \frac{%
d^{4}q}{(2\pi )^{4}}A_{\mu }^{\alpha }(q)\mathcal{P}_{\mu \nu }^{\alpha
\beta }A_{\nu }^{\beta }(-q)-\frac{1}{2}\int \frac{d^{4}q}{(2\pi )^{4}}%
A_{\mu }^{3}(q)\mathcal{Q}_{\mu \nu }A_{\nu }^{3}(-q)},  \label{Zq1}
\end{equation}
with
\begin{eqnarray}
\mathcal{P}_{\mu \nu }^{\alpha \beta } &=&\delta ^{\alpha \beta }\left(
\delta _{\mu \nu }\left( q^{2}+\nu ^{2}g^{2}\right) +\left( \frac{1}{\xi }%
-1\right) q_{\mu }q_{\nu }+\frac{g^{2}}{2}\left( \beta +\frac{\omega }{2}%
\right) \frac{1}{q^{2}}\delta _{\mu \nu }\right)  \label{P}  \;, \nonumber \\
\mathcal{Q}_{\mu \nu } &=&\delta _{\mu \nu }\left( q^{2}+\frac{\omega g^{2}%
}{2}\frac{1}{q^{2}}\right) +\left( \frac{1}{\xi }-1\right) q_{\mu }q_{\nu } \;,
\label{Q}
\end{eqnarray}
where again $\xi\to0$ is understood to recover the
Landau gauge. Evaluating  the inverse of the
expressions above and taking the limit $\xi\rightarrow 0$, for the gluon propagators one gets
\begin{eqnarray}
\left\langle A_{\mu }^{3}(q)A_{\nu }^{3}(-q)\right\rangle &=&\frac{q^{2}}{%
q^{4}+\frac{\omega g^{2}}{2}}\left( \delta _{\mu \nu }-\frac{q_{\mu }q_{\nu
}}{q^{2}}\right)  \label{Pdiag} \;, \\
\left\langle A_{\mu }^{\alpha }(q)A_{\nu }^{\beta }(-q)\right\rangle
&=&\delta ^{\alpha \beta }\frac{q^{2}}{q^{2}\left( q^{2}+g^{2}\nu^{2}\right)
+g^{2}\left( \frac{\beta }{2}+\frac{\omega }{4}\right) }\left( \delta _{\mu
\nu }-\frac{q_{\mu }q_{\nu }}{q^{2}}\right)  \label{NPoff} \;.
\end{eqnarray}
To establish the gap equations for the Gribov parameters $(\beta, \omega)$, we evaluate the partition function $Z_{quad}$ in the semiclassical approximation. As done before, we integrate out the gauge fields, obtaining
\begin{equation}
Z_{quad}=\int{\frac{d\beta}{2\pi i\beta}\frac{d\omega}{2\pi i\omega}}%
e^{\beta}e^{\omega}\left(\det\mathcal{Q}_{\mu\nu}\right)^{-\frac{1}{2}%
}\left(\det\mathcal{P}^{\alpha\beta}_{\mu\nu}\right)^{-\frac{1}{2}} \;.
\label{Zq2}
\end{equation}
where the determinants in expression (\ref{Zq2}) are given by
\begin{eqnarray}
\left( \det \mathcal{Q}_{\mu \nu }\right) ^{-\frac{1}{2}} &=&\exp \left[
-\frac{3}{2}\int {\frac{d^{4}q}{(2\pi )^{4}}\ln \left( q^{2}+\frac{\omega g^{2}}{2}%
\frac{1}{q^{2}}\right) }\right] \;, \nonumber \\
\left( \det \mathcal{P}_{\mu \nu }^{\alpha \beta }\right) ^{-\frac{1}{2}}
&=&\exp \left[ -3\int {\frac{d^{4}q}{(2\pi )^{4}}\ln \left(
(q^{2}+g^{2}\nu^{2})+g^{2}\left( \frac{\beta }{2}+\frac{\omega }{4}\right)
\frac{1}{q^{2}}\right) }\right] \;.
\end{eqnarray}
Therefore,
\begin{equation}  \label{Zq3}
Z_{quad}=\int{\frac{d\beta}{2\pi i}\frac{d\omega}{2\pi i}}e^{f(\omega,\beta)} \;,
\end{equation}
where, in the thermodynamic limit \cite{Gribov:1977wm,Sobreiro:2005ec,Vandersickel:2012tz}
\begin{eqnarray}
f(\omega ,\beta ) &=&\beta +\omega  -\frac{3}{2}\int {\frac{d^{4}q%
}{(2\pi )^{4}}\ln \left( q^{2}+\frac{\omega g^{2}}{2}\frac{1}{q^{2}}\right)
} \nonumber  \\
&-&3\int {\frac{d^{4}q}{(2\pi )^{4}}\ln \left(
(q^{2}+g^{2}\nu^{2})+g^{2}\left( \frac{\beta }{2}+\frac{\omega }{4}\right)
\frac{1}{q^{2}}\right) }.  \label{fbeta}
\end{eqnarray}
We again proceed by evaluating expression (\ref{Zq3})  in the saddle point approximation \cite{Gribov:1977wm,Sobreiro:2005ec,Vandersickel:2012tz}, {\it i.e.}
\begin{equation}
Z_{quad}\approx e^{f(\beta^*,\omega^*)} \;,
\end{equation}
where the parameters $\beta^*$ and $\omega^*$ are determined by the stationary conditions
\begin{equation}
\frac{\partial f}{\partial \beta^*}=\frac{\partial f}{\partial \omega^*}=0 \;,
\end{equation}
which yield the following gap equations:
\begin{eqnarray}
\frac{3}{2}\left( \frac{g^{2}}{2}\right) \int \frac{d^{4}q}{(2\pi )^{4}}\left( \frac{1}{q^{4}+\frac{%
\omega ^{\ast }g^{2}}{2}}+\frac{1}{q^{2}(q^{2}+g^{2}\nu^{2})+g^{2}\left(
\frac{\beta ^{\ast }}{2}+\frac{\omega ^{\ast }}{4}\right) }\right) &=&1 \;, \label{gap1} \\
3 \left(\frac{g^{2}}{2}\right) \int \frac{d^{4}q}{(2\pi )^{4}}\left( \frac{1}{%
q^{2}(q^{2}+g^{2}\nu^{2})+g^{2}\left( \frac{\beta ^{\ast }}{2}+\frac{\omega
^{\ast }}{4}\right) }\right) &=&1 \;, \label{gap2}
\end{eqnarray}
allowing us to express $\beta^*, \omega^*$ in terms of the parameters $\nu,g$.

Let us start by considering the second gap equation, eq.\eqref{gap2}, and decompose the denominator according to
\begin{equation}
q^4 + g^2 \nu^2 q^2 + \tau = (q^2+q^2_+) (q^2+q^2_-) \;, \qquad \tau = g^2 \left( \frac{\beta^*}{2}+\frac{\omega^*}{4} \right)  \;, \label{dec}
\end{equation}
with
\begin{equation}
q^2_+ = \frac{1}{2} \left(g^2 \nu^2 + \sqrt{g^4\nu^4 -4 \tau}   \right) \;,  \qquad q^2_- = \frac{1}{2} \left(g^2 \nu^2 - \sqrt{g^4\nu^4 -4 \tau}   \right) \;.
\label{roots}
\end{equation}
Let us discuss first the case of two real,  positive, different roots, namely $0 \le \tau <  \frac{g^4\nu^4}{4}$. From
\eqref{intd2f}, we reexpress the gap equation \eqref{gap2} as
\begin{equation}
\left(1 + \frac{q^2_-}{q^2_+ -  q^2_-}\; \ln\left( \frac{q^2_-}{{\omu}^2} \right)  - \frac{q^2_+}{q^2_+ -  q^2_-}\; \ln\left( \frac{q^2_+}{{\omu}^2} \right)  \right) = \frac{32\pi^2}{3g^2}  \;. \label{gap2bb}
\end{equation}
Introducing now the dimensionless variables\footnote{We introduced the renormalization group invariant scale $\lms$. }
\begin{eqnarray}
b & = &  \frac{g^2\nu^2}{2 {\bar \mu}^2\; e^{\left(1-\frac{32\pi^2}{3g^2}\right)} } =\frac{1}{2\; e^{\left( 1 -\frac{272 \pi^2}{21 g^2} \right)}}  \;\frac{g^2\nu^2}{\lms^2  } \;,   \nonumber \\[3mm]
 \xi  &=  &\frac{4\tau}{g^4\nu^4}  \geq 0\;,  \qquad
0 \le \xi < 1 \;, \label{vb}
\end{eqnarray}
and proceeding as in the previous case, equation  \eqref{gap2bb} can be recast in the following form
\begin{equation}
2 \sqrt{1-\xi}\; \ln(b) = -  \left( 1 +  \sqrt{1-\xi} \right) \; \ln\left( 1 +  \sqrt{1-\xi} \right) +  \left( 1 -  \sqrt{1-\xi} \right) \; \ln\left( 1 - \sqrt{1-\xi} \right)  \;. \label{gapdd}
\end{equation}
or compactly,
\begin{equation}
2\ln b=g(\xi) \label{gapddbis}
\end{equation}
with the same function as already defined in the fundamental sector, eqns.~\eqref{g},\eqref{gex}.  Also here, the equation \eqref{gapddbis} remains valid also for complex conjugate roots, viz.~$\xi>1$. We are then easily lead to the following cases.

\subsection{The case $b<\frac{1}{2}$}
Using the properties of $g(\xi)$, it turns out that eq.\eqref{gapdd} admits a unique solution for $\xi$, which can be explicitly constructed with a numerical approach. More precisely, when the mass scale $g^2\nu^2$ is sufficiently smaller  than $\lms^2$, {\it i.e.}
\begin{equation}
g^2 \nu^2 < 2 \; e^{\left( 1 -\frac{272 \pi^2}{21 g^2} \right)} \; \lms^2   \;, \label{mh}
\end{equation}
we have what can be called a $U(1)$ confined phase. In fact, in this case, the gap equations \eqref{gap1} and \eqref{gap2} can be rewritten as
\begin{eqnarray}
3 \left( \frac{g^{2}}{2}\right) \int \frac{d^{4}q}{(2\pi )^{4}}\left( \frac{1}{%
q^{4}+g^2 \frac{\omega^*}{2}}\right) &=&1\;, \label{ggapp1} \\
3 \left( \frac{g^{2}}{2}\right) \int \frac{d^{4}q}{(2\pi )^{4}}\left( \frac{1%
}{q^{2}(q^{2}+g^{2}\nu^{2})+g^2 (\frac{\beta^*}{2}+\frac{\omega^*}{4})}
\right) &=&1\label{ggapp2} \;.
\end{eqnarray}
Moreover, making use of
\begin{equation}
\int \frac{d^dp}{(2\pi)^d} \frac{1}{p^4 + m^4}  = \frac{2}{\bar \varepsilon} \frac{1}{16\pi^2} - \frac{1}{16\pi^2} \left( \ln\frac{m^2}{\omu^2} - 1 \right) \;,  \label{intd1}
\end{equation}
the gap equation \eqref{ggapp1} gives
\begin{equation}
{\left( \frac{g^2 \omega^*}{2} \right)}^{1/2} = {\bar \mu}^2  e^{ \left(1 - \frac{32\pi^2}{3g^2} \right)}  = \lms^2\; e^{\left( 1 -\frac{272 \pi^2}{21 g^2} \right)}  \;.
\end{equation}
Therefore, for $b<\frac{1}{2}$, the $A^3_\mu$ component of the gauge field gets confined, exhibiting a Gribov-type propagator with complex poles, namely
\begin{equation}
\left\langle A_{\mu }^{3}(q)A_{\nu }^{3}(-q)\right\rangle =\frac{q^{2}}{%
q^{4}+\frac{\omega^* g^{2}}{2}}\left( \delta _{\mu \nu }-\frac{q_{\mu }q_{\nu
}}{q^{2}}\right)  \;. \label{Pdiag_conf}
\end{equation}
Relying on eq.\eqref{gapdd} we can then distinguish the two phases:
\begin{itemize}
\item[(i)]  when $\frac{1}{e}<b<\frac{1}{2}$, equation  \eqref{gapdd} has a single solution with $0 \le \xi <1$. In this region, the roots $(q^2_+, q^2_-)$  are thus real and  the off-diagonal propagator decomposes into the sum of two Yukawa propagators
\begin{equation}
\left\langle A_{\mu }^{\alpha }(q)A_{\nu }^{\beta }(-q)\right\rangle
=\delta ^{\alpha \beta } \left(  \frac{{\cal R}_+}{q^2+q^2_+} -    \frac{{\cal R}_-}{q^2+q^2_-}   \right)
\left( \delta _{\mu
\nu }-\frac{q_{\mu }q_{\nu }}{q^{2}}\right)  \label{NPoff_fin} \;,
\end{equation}
where
\begin{equation}
{\cal R}_+ = \frac{q^2_+}{q^2_+-q^2_-}  \;, \qquad {\cal R}_- = \frac{q^2_-}{q^2_+-q^2_-} \;. \label{r}
\end{equation}
However, due to the relative minus sign in eq.\eqref{NPoff_fin},  only the component $ {\cal R}_+ $ can be associated to a physical mode, analogously as in the fundamental case. Also here, a further study using BRST tools will be devoted to this. Due to the confinement of the third component $A^3_\mu$, this phase is referred as the $U(1)$ confining phase. It is worth observing that it is also present in the $3d$ case, with terminology coined in \cite{Nadkarni:1989na}, see also \cite{Capri:2012cr}.
\item[(ii)]  for $b<\frac{1}{e}$, equation \eqref{gapdd} has a solution for  $\xi>1$. In this region the roots  $(q^2_+, q^2_-)$  become complex conjugate and the off-diagonal gluon propagator is of the Gribov type, displaying complex poles.  In this region all gauge fields display a propagator of the Gribov type. This is the $SU(2)$ confining region.
\end{itemize}
Similarly, the above regions are continuously connected when $b$ varies. In particular, for $b \stackrel{<}{\to} \frac{1}{2}$, we obtain $\xi=0$ as solution.

\subsection{The case $b>\frac{1}{2}$}
Let us consider now the case in which $b>\frac{1}{2}$. Here, there is no solution of the equation \eqref{gapdd} for the parameter $\xi$, as it follows by observing that the left hand side of eq.\eqref{gapdd} is always positive, while the right hand side is always negative. This has a deep physical consequence. It means that for a Higgs mass $m_{Higgs}^2 = g^2\nu^2$ sufficiently larger than $\lms^2$, {\it i.e.}
\begin{equation}
g^2 \nu^2 > 2 \; e^{\left( 1 -\frac{272 \pi^2}{21 g^2} \right)} \; \lms^2   \;, \label{mh}
\end{equation}
the gap equation \eqref{gap2} is inconsistent. It is then important to realize that this is actually the gap equation obtained by acting with $\frac{\p}{\p \beta}$ on the vacuum energy  ${\cal E}_v= -  f(\omega ,\beta )$, eq.\eqref{fbeta}.
So, we are forced to set $\beta^*=0$, and confront the remaining $\omega$-equation, viz.~eq.\eqref{gap1}:
\begin{equation}
\frac{3}{2}\left( \frac{g^{2}}{2}\right) \int \frac{d^{4}q}{(2\pi )^{4}}\left( \frac{1}{q^{4}+\frac{%
\omega ^{\ast }g^{2}}{2}}+\frac{1}{q^{2}(q^{2}+g^{2}\nu^{2})+\frac{g^2\omega ^{\ast }}{4} }\right) =1 \;, \label{gap1n}
\end{equation}
which can be transformed into
\begin{eqnarray}
4 \; \ln(b) & = &  \frac{1}{\sqrt{1-\xi}} \left[ -\left( 1 +  \sqrt{1-\xi} \right) \; \ln\left( 1 +  \sqrt{1-\xi} \right) +  \left( 1 -  \sqrt{1-\xi} \right) \; \ln\left( 1 - \sqrt{1-\xi} \right) \right.  \nonumber \\
 &\; \;&-\left. \sqrt{1-\xi}\; \ln\xi - \sqrt{1-\xi} \ln 2\right]\equiv h(\xi)
 \label{gapddn}
\end{eqnarray}
after a little algebra, where $\xi=\frac{\omega^*}{g^4\nu^4}$. The behaviour of $h(\xi)$ for $\xi\geq0$ is more complicated than that of $g(\xi)$. Because of the $-\ln\xi$ contribution, $h(\xi)$ becomes more and more positive when $\xi$ approaches zero. In fact, $h(\xi)$ strictly decreases from $+\infty$ to $-\infty$ for $\xi$ ranging from 0 to $+\infty$.

It is interesting to consider first the limiting case $b \stackrel{>}{\to} \frac{1}{2}$, yielding $\xi\approx 1.0612$.   So, there is a discontinuous jump in $\xi$ (i.e.~the Gribov parameter for fixed $v$) when the parameter $b$ crosses the boundary value $\frac{1}{2}$.

We were able to separate the $b>\frac{1}{2}$ region as follows:
\begin{itemize}
\item[(a)] For $\frac{1}{2}<b<\frac{1}{\sqrt{\sqrt{2}e}}\approx 0.51$, we have a unique solution $\xi>1$, i.e.~we are in the confining region again, with all gauge bosons displaying a Gribov type of propagator with complex conjugate poles.
\item[(b)] For $\frac{1}{\sqrt{\sqrt{2}e}}< b<\infty$, we have a unique solution $\xi<1$, indicating again a combination of two Yukawa modes for the off-diagonal gauge bosons. The ``photon'' is still of the Gribov type, thus confined.
\end{itemize}
Completely analogous as in the fundamental case, it can be checked by addressing the averages of the expressions \eqref{sigma} that for $b>\frac{1}{2}$ and $\omega$ obeying the gap equation with $\beta=0$, we are already within the Gribov horizon, making the introduction of the second Gribov parameter $\beta$ obsolete.

It is obvious that the transitions in the adjoint case are far more intricate than in the earlier studied fundamental case. First of all, we notice that the ``photon'' (diagonal gauge boson) is confined according to its Gribov propagator. There is never a Coulomb phase for $b<\infty$. The latter finding is can be understood again from the viewpoint of the ghost self-energy. If the diagonal gluon would remain Coulomb (massless), the off-diagonal ghost self-energy, cfr.~eq.\eqref{sigma}, will contain an untamed infrared contribution from this massless photon\footnote{The ``photon'' indeed keeps it coupling to the charged (= off-diagonal) ghosts, as can be read off directly from the Faddeev-Popov term $c^a \p_\mu D_\mu^{ab}c^b$.}, leading to an off-diagonal ghost self-energy that will cross the value 1 at a momentum $k^2>0$, indicative of trespassing the first Gribov horizon. This crossing will not be prevented at any finite value of the Higgs condensate $\nu$, thus we are forced to impose at any time a nonvanishing Gribov parameter $\omega$. Treating the gauge copy problem for the adjoint Higgs sector will screen (rather confine) the a priori massless ``photon''.

An interesting limiting case is that of infinite Higgs condensate, also considered in the lattice study of \cite{Brower:1982yn}. Assuming $\nu\to\infty$, we have $b\to\infty$ according to its definition \eqref{vb}. Expanding the gap equation \eqref{gapddn} around $\xi=0^+$, we find the limiting equation $b^4=\frac{1}{\xi}$, or equivalently $\omega^*\propto \lms^8/g^4\nu^4$. Said otherwise, we find that also the second Gribov parameter vanishes in the limit of infinite Higgs condensate. As a consequence, the photon becomes truly massless in this limit. This result provides ---in our opinion--- a kind of continuum version of the existence of the Coulomb phase in the same limit as in the lattice version of the model probed in \cite{Brower:1982yn}. It is instructive to link this back to the off-diagonal no pole function, see Eq.~\eqref{sigma}, as we have argued in the proceeding paragraph that the massless photon leads to $\sigma_{off}(0)>1$ upon taking averages. However, there is an intricate combination of the limits $\nu\to\infty$, $\omega^*\to 0$ preventing such a problem here. Indeed, we find in these limits, again using dimensional regularization in the $\MSbar$ scheme, that
\begin{align}\label{dv}
  \sigma_{off}(0) &= \frac{3g^2}4 \left(\int \frac{d^4q}{(2\pi)^4} \frac1{q^4+\frac{\omega^\ast g^2}2} + \int \frac{d^4q}{(2\pi)^4} \frac1{q^2(q^2+g^2\nu^2)+\frac{\omega^\ast g^2}4}\right) \nonumber\\
	&= -\frac{3g^2}{128\pi^2} \left(\tfrac12 \ln\frac{\omega^\ast g^2}{2\overline\mu^4} + \ln\frac{g^2\nu^2}{\overline\mu^2}-2\right) \nonumber \\
	&= -\frac{3g^2}{128\pi^2} \left(\tfrac12 \ln\frac{\omega^\ast g^6\nu^4}{2\overline\mu^8}-2\right) \stackrel{b^4=\xi^{-1}}{\longrightarrow} -\frac{3g^2}{64\pi^2} \ln 8g^2 + \frac12.
\end{align}
The latter quantity is always smaller than $1$ for $g^2$ positive, meaning that we did not cross the Gribov horizon. This observation confirm in an explicit way the intuitive reasoning also found in section 3.4 of \cite{Lenz:2000zt}, at least in the limit $\nu\to\infty$. The subtle point in the above analysis is that it is not allowed to naively throw away the 2nd integral in the first line of \eqref{dv} for $\nu\to\infty$. There is a logarithmic $\ln\nu$ ($\nu\to\infty$) divergence that conspires with the $\ln \omega^*$ ($\omega^*\to0$) divergence of the 1st integral to yield the final reported result. This displays that, as usual, certain care is needed when taking infinite mass limits in Feynman integrals.

\subsection{The vacuum energy in the adjoint case}
As done in the case of the fundamental representation, let us work out the expression of the vacuum energy ${\cal E}_v$, for which we have the one loop  integral representation
\begin{eqnarray}
{\cal E}_v= -\beta^*-\omega^*+\frac{3}{2}\int \frac{d^4q}{(2\pi)^4}\ln(q^4+2\tau')+3\int \frac{d^4q}{(2\pi)^4} \ln(q^4+g^2\nu^2q^2+\tau)
\end{eqnarray}
where
\begin{equation}\label{ddvac1}
    \tau'=\frac{g^2\omega^*}{4}\,,\qquad \tau=g^2\left(\frac{\beta^*}{2}+\frac{\omega^*}{4}\right).
\end{equation}
Based on \eqref{intdl} and on
\begin{equation}\label{ddvac2}
    \int\frac{d^4q}{(2\pi)^4}\ln(q^4+m^4)=-\frac{m^4}{32\pi^2}\left(\ln\frac{m^4}{\omu^4}-3\right),
\end{equation}
we find
\begin{eqnarray}\label{ddvac3}
{\cal E}_v= -\frac{2}{g^2}(\tau+\tau')-\frac{3\tau'}{32\pi^2}\left(\ln\frac{2\tau'}{\omu^4}-3\right)+\frac{3}{32\pi^2}\left(q_-^4\left(\ln\frac{q_-^2}{\omu^2}-\frac{3}{2}\right)+q_+^4\left(\ln\frac{q_+^2}{\omu^2}-\frac{3}{2}\right)\right).
\end{eqnarray}
We can introduce $b$ via its definition \eqref{vb} to write after simplification
\begin{eqnarray}\label{ddvac4}
\frac{{\cal E}_v}{g^4\nu^4}&=& -\frac{1}{g^2}-\frac{3\xi'}{128\pi^2}\left(\ln(2b^2\xi')-1\right)+\frac{3(4-2\xi)}{128\pi^2}\left(\ln b-\frac{1}{2}\right)\nonumber\\
&&+\frac{3}{128\pi^2}\left(\left(1-\sqrt{1-\xi}\right)^2\ln\left(1-\sqrt{1-\xi}\right)+\left(1+\sqrt{1-\xi}\right)^2\ln\left(1+\sqrt{1-\xi}\right)\right)\;,
\end{eqnarray}
with
\begin{equation}\label{ddvac5}
    \xi'=\frac{4\tau'}{g^4\nu^4}\,,\qquad \xi=\frac{4\tau}{g^4\nu^4} \;.
\end{equation}
First, for $b<1/2$, we can use the  $\omega$-gap equation \eqref{ggapp1} to establish $\xi'=\frac{1}{2b^2}$, and thus
\begin{eqnarray}\label{ddvac4bis}
\frac{{\cal E}_v}{g^4\nu^4}&=& -\frac{1}{g^2}+\frac{3}{256b^2\pi^2}+\frac{3(4-2\xi)}{128\pi^2}\left(\ln b-\frac{1}{2}\right)\nonumber\\
&&+\frac{3}{128\pi^2}\left(\left(1-\sqrt{1-\xi}\right)^2\ln\left(1-\sqrt{1-\xi}\right)+\left(1+\sqrt{1-\xi}\right)^2\ln\left(1+\sqrt{1-\xi}\right)\right)  \;,
\end{eqnarray}
where $\xi$ is determined by the equations  \eqref{gapdd},\eqref{gapddbis}.

For $b>1/2$, we can remember that $\beta=0$ and thus $\xi=\xi'$, in which case we can easily obtain
\begin{eqnarray}
\frac{{\cal E}_v}{g^4 \nu^4} & = & - \left( \frac{1}{g^2} + \frac{3}{64\pi^2}  \right)   + \frac{3}{64\pi^2} \xi+ \frac{3}{32\pi^2} \frac{1}{4} (4-2\xi) \ln(b) \nonumber \\
&+& \frac{3}{32\pi^2} \frac{1}{4} \left( \left(1+\sqrt{1-\xi}\right)^2\;\ln\left(1+\sqrt{1-\xi}\right) + \left(1-\sqrt{1-\xi}\right)^2\;\ln\left(1-\sqrt{1-\xi}\right) \right) \nonumber \\
&-& \frac{3}{32\pi^2} \frac{1}{4} \left(2\xi \ln(\sqrt{\xi}) +2\xi \ln\sqrt{2} + 2\xi \ln(b)   \right)   \;, \label{Evv2}
\end{eqnarray}
with $\xi$ now given by eq.\eqref{gapddn}.

\begin{figure}[h!]
\center
\includegraphics[width=8cm]{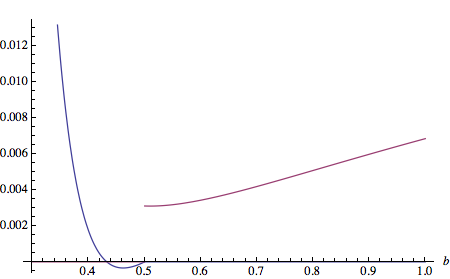}
\caption{Plot of the vacuum energy in the adjoint representation as a function of the parameter $b$. The discontinuity at $b=\frac{1}{2}$ is evident. }
\label{BS-d4}
\end{figure}

It is worth noticing that the discontinuity in the parameter $\xi$ directly reflects itself in a discontinuity in the vacuum energy, as is clear from the plot of Fig.\ref{BS-d4}.

Investigating the functional \eqref{ddvac4} in terms of $\xi$ and $\xi'$, it is numerically (graphically) rapidly established there is always a solution to the gap equations $\frac{\p \mathcal{E}_v }{\p \xi}=\frac{\p \mathcal{E}_v }{\p \xi'}=0$ for $b<\frac{1}{2}$, but the solution $\xi^*$ is pushed towards the boundary $\xi=0$ if $b$ approaches $\frac{1}{2}$, to subsequently disappear for $b>\frac{1}{2}$ \footnote{The gap solutions correspond to a local maximum, as identified by analyzing the Hessian matrix of 2nd derivatives.}. In that case, we are forced to return on our steps as in the fundamental case and conclude that $\beta=0$, leaving us with a single variable $\xi=\xi'$ and a new vacuum functional to extremize. There is a priori no reason why these 2 intrinsically different vacuum functionals would be smoothly joined at $b=\frac{1}{2}$. This situation is clearly different from what happens when a potential has e.g.~2 different local minima with different energy, where at a first order transition the two minima both become global minima, thereafter changing their role of local vs.~global. Evidently, the vacuum energy does not jump since it is by definition equal at the transition.

Nevertheless, a completely analogous analysis as for the fundamental case will learn that $b=\frac{1}{2}$ is beyond the range of validity of our approximation\footnote{A little more care is needed as the appearance of two Gribov scales complicate the log structure. However, for small $b$ the Gribov masses will dominate over the Higgs condensate and we can take  $\omu$ of the order of the Gribov masses to control the logs and get a small coupling. For large $b$, we have $\beta^*=0$ and a small $\omega^*$: the first log will be kept small by its prefactor and the other logs can be managed by taking $\omu$ of the order of the Higgs condensate.}. The small and large $b$ results can again be shown to be valid, so at large $b$ ($\sim$ large Higgs condensate) we have a mixture of off-diagonal Yukawa and confined diagonal modes and at small $b$ ($\sim$ small Higgs condensate) we are in a confined phase. In any case we have that the diagonal gauge boson is \emph{not} Coulomb-like, its infrared behaviour is suppressed as it feels the presence of the Gribov horizon.

\section{Conclusion}

In this work we have attempted at studying the transition between the Higgs and the confinement phase within a continuum quantum field theory. The problem has been addressed by restricting the domain of integration in the functional integral to the so called Gribov region $\Omega$, which enables us to take into account the nonperturbative effect of the Gribov copies. This framework allows  us to discuss the transition between the Higgs phase and the confinement phase by looking at the pole structure of the two-point gluon correlation function. Both fundamental and adjoint representation for the Higgs field have been considered. The output of our investigation reveals that the case of the fundamental representation is different from that of the adjoint representation, a feature in agreement with the results of numerical lattice simulations \cite{Fradkin:1978dv,Lang:1981qg,Langguth:1985dr,Azcoiti:1987ua,Caudy:2007sf,Bonati:2009pf,Maas:2010nc,Maas:2012zf,Greensite:2011zz}.

In the case of the fundamental representation, the gluon propagator evolves in a continuous way from a confining propagator of the Gribov type to a Yukawa type propagator describing the Higgs phase. Again, this feature is in qualitative agreement with lattice results  \cite{Fradkin:1978dv,Lang:1981qg,Langguth:1985dr,Azcoiti:1987ua,Caudy:2007sf,Bonati:2009pf,Maas:2010nc,Maas:2012zf,Greensite:2011zz}, which show that the transition between the Higgs and the confining phase occurs in a continuous way.  Moreover, we have been able to show that, in the weak coupling region, {\it i.e.} in the Higgs phase, there is no need to implement the restriction to the Gribov region $\Omega$. Said otherwise, in this region, the presence of the Higgs field automatically ensures that the theory lies within the Gribov region, so that the Gribov horizon is not crossed. This is a relevant result, implying that, at weak coupling, the usual Higgs mechanism takes place, being not affected by the existence of the Gribov copies. A safely asymptotic non-confining theory can be introduced, with massive gauge bosons as asymptotic states.

In the adjoint representation things look quite different. Besides the confining phase, in which the gluon propagator is of the Gribov type, our results indicate the existence of what can be called a $U(1)$ confining phase for finite values of the Higgs condensate. This is a phase in which the third component $A^3_\mu$ of the gauge field  displays a propagator of the Gribov type, while the remaining off-diagonal components $A^{\alpha}_\mu$, $\alpha=1,2$, exhibit a propagator of the Yukawa type. Interestingly, this phase has been already detected in the lattice studies of the three dimensional Georgi-Glashow model   \cite{Nadkarni:1989na,Hart:1996ac}. A second result of our analysis is the absence of the Coulomb phase for finite Higgs condensate. For an infinite value of the latter, we were able to clearly reveal the existence of a massless photon, in agreement with the lattice suggestion of \cite{Brower:1982yn}.

Summarizing, it seems safe to state that the results we have obtained so far can be regarded as being in qualitative agreement with the lattice findings and worth to be pursued. Let us end by giving a preliminary list of points for future  investigation:

\begin{itemize}

\item in the pure gauge case, it has been shown in past years that the Gribov theory dynamically corrects itself via the condensation of the auxiliary fields arising when the restriction to the Gribov region is implemented in a local and renormalizable way. This has led to the Refined Gribov-Zwanziger (RGZ) theory   \cite{Dudal:2007cw,Dudal:2008sp,Dudal:2011gd,Gracey:2010cg}, resulting in propagators and dynamics in very good agreement with lattice investigations  \cite{Dudal:2010tf,Cucchieri:2011ig,Oliveira:2012eh,Dudal:2012zx}. It would be worth to discuss the transition between the Higgs and confining phase within the RGZ framework. This might give more reliable quantitative results to be compared with the lattice data.

\item within the present approximation,  we were not able to address in a concrete and detailed way the issue of the characterization of the phase diagram in the $(\nu,g)$ plane. To that aim, it might  be interesting to embed in our framework the Polyakov loop $\mathcal{P}$ as  order parameter of the Higgs-confinement transition, as  studied in \cite{Bonati:2009pf}. In \cite{Fukushima:2012qa}, the pure gauge thermodynamics at finite $T$ was investigated by  using the RGZ gluon and ghost propagators. In particular, an approximate effective action for $\left\langle \mathcal{P}\right\rangle$ was constructed. We can try to couple $\left\langle \mathcal{P}\right\rangle$ to the Gribov effective action and investigate the interplay of both Gribov mass and $\left\langle \mathcal{P}\right\rangle$ in terms of a varying Higgs condensate $g^2\nu^2$, the latter quantity playing the role of temperature as in \cite{Fukushima:2012qa}.

\item certainly,  the extension of the present investigation to the case of the gauge group $SU(2) \times U(1)$ \cite{prep} has an apparent interest, due to its relationship with the electroweak theory. Since the latter is based on a fundamental Higgs and since we have shown in this paper that in the ``perturbative regime'' (sufficiently large Higgs condensate and small effective coupling constant) in the fundamental case the Gribov dynamics becomes trivial, we can expect to recover at least a massless photon for the QED part of the $SU(2) \times U(1)$ gauge theory in the same region of (physically relevant) parameter space. Our preliminary findings \cite{prep} do support this naive extrapolation of the here presented work. The concrete analysis is however rather cumbersome, details are thus to be reported at a later stage in \cite{prep}.

\item finally, we hope that our results will motivate further lattice investigations on the behavior of the gluon and ghost propagators  in presence of Higgs fields. Although these studies have been already started in the case of the fundamental representation \cite{Maas:2010nc,Maas:2012zf}, it would be quite interesting to have at our disposal also data for the case of the adjoint representation.

\end{itemize}

\section*{Acknowledgments}
The Conselho Nacional de Desenvolvimento Cient\'{\i}fico e
Tecnol\'{o}gico (CNPq-Brazil), the Faperj, Funda{\c{c}}{\~{a}}o de
Amparo {\`{a}} Pesquisa do Estado do Rio de Janeiro, the Latin
American Center for Physics (CLAF), the SR2-UERJ,  the
Coordena{\c{c}}{\~{a}}o de Aperfei{\c{c}}oamento de Pessoal de
N{\'{\i}}vel Superior (CAPES)  are gratefully acknowledged. D.~D.~is supported by the Research-Foundation Flanders. We thank \v{S}.~Olejn\'{i}k for discussions.


\begin{thebibliography}{99}




\bibitem{Fradkin:1978dv}
  E.~H.~Fradkin and S.~H.~Shenker,
  Phys.\ Rev.\ D {\bf 19} (1979) 3682.

\bibitem{Lang:1981qg}
C.~B.~Lang, C.~Rebbi and M.~Virasoro,
  Phys.\ Lett.\ B {\bf 104} (1981) 294.

\bibitem{Langguth:1985dr}
W.~Langguth, I.~Montvay and P.~Weisz,
  Nucl.\ Phys.\ B {\bf 277} (1986) 11.

\bibitem{Azcoiti:1987ua}
V.~Azcoiti, G.~Di Carlo, A.~F.~Grillo, A.~Cruz and A.~Tarancon,
  Phys.\ Lett.\ B {\bf 200} (1988) 529.

\bibitem{Caudy:2007sf}
  W.~Caudy and J.~Greensite,
  Phys.\ Rev.\ D {\bf 78} (2008) 025018
  [arXiv:0712.0999 [hep-lat]].

\bibitem{Bonati:2009pf}
  C.~Bonati, G.~Cossu, M.~D'Elia and A.~Di Giacomo,
  Nucl.\ Phys.\ B {\bf 828} (2010) 390
  [arXiv:0911.1721 [hep-lat]].

\bibitem{Maas:2010nc}
  A.~Maas,
  Eur.\ Phys.\ J.\ C {\bf 71} (2011) 1548
  [arXiv:1007.0729 [hep-lat]].

\bibitem{Maas:2012zf}
A.~Maas and T.~Mufti,
  arXiv:1211.5301 [hep-lat].


\bibitem{Greensite:2011zz}
  J.~Greensite,
  Lect.\ Notes Phys.\  {\bf 821} (2011) 1.


\bibitem{Horowitz:1983sr}
A.~M.~Horowitz,
  Nucl.\ Phys.\ B {\bf 235} (1984) 563.

\bibitem{Damgaard:1985nb}
P.~H.~Damgaard and U.~M.~Heller,
  Phys.\ Lett.\ B {\bf 164} (1985) 121.

\bibitem{Baier:1986sa}
R.~Baier and H.~J.~Reusch,
  Nucl.\ Phys.\ B {\bf 285} (1987) 535.

\bibitem{Fister:2010yw}
L.~Fister, R.~Alkofer and K.~Schwenzer,
  Phys.\ Lett.\ B {\bf 688} (2010) 237
  [arXiv:1003.1668 [hep-th]].

\bibitem{Macher:2011ys}
V.~Macher, A.~Maas and R.~Alkofer,
  Int.\ J.\ Mod.\ Phys.\ A {\bf 27} (2012) 1250098
  [arXiv:1106.5381 [hep-ph]].

\bibitem{Osterwalder:1977pc}
 K.~Osterwalder and E.~Seiler,
  Annals Phys.\  {\bf 110} (1978) 440.

\bibitem{leeyang}
T.~D.~Lee and C.~N.~Yang, Phys.\ Rev.\ {\bf 87} (1952) 410.


\bibitem{Nussinov:2004ns}
Z.~Nussinov,
  Phys.\ Rev.\ D {\bf 72} (2005) 054509.
  [cond-mat/0411163].



\bibitem{Gribov:1977wm} V.~N.~Gribov,
Nucl.\ Phys.\ B \textbf{139} (1978) 1. 

\bibitem{Sobreiro:2005ec}
R.~F.~Sobreiro and S.~P.~Sorella,
arXiv:hep-th/0504095. 

\bibitem{Vandersickel:2012tz}
  N.~Vandersickel and D.~Zwanziger,
  Phys.\ Rept.\  {\bf 520} (2012) 175
  [arXiv:1202.1491 [hep-th]].

\bibitem{Capri:2012cr}
  M.~A.~L.~Capri, D.~Dudal, A.~J.~Gomez, M.~S.~Guimaraes, I.~F.~Justo and S.~P.~Sorella,
  arXiv:1210.4734 [hep-th].


  \bibitem{Nadkarni:1989na}
  S.~Nadkarni,
  Nucl.\ Phys.\ B {\bf 334} (1990) 559.

\bibitem{Hart:1996ac}
  A.~Hart, O.~Philipsen, J.~D.~Stack and M.~Teper,
  Phys.\ Lett.\ B {\bf 396} (1997) 217
  [hep-lat/9612021].

\bibitem{Brower:1982yn}
R.~C.~Brower, D.~A.~Kessler, T.~Schalk, H.~Levine and M.~Nauenberg,
  Phys.\ Rev.\ D {\bf 25} (1982) 3319.


\bibitem{Greensite:2004ke}
 J.~Greensite, S.~Olejnik and D.~Zwanziger,
  Phys.\ Rev.\ D {\bf 69} (2004) 074506
  [hep-lat/0401003].

\bibitem{Dudal:2010fq}
D.~Dudal, S.~P.~Sorella and N.~Vandersickel,
  Eur.\ Phys.\ J.\ C {\bf 68} (2010) 283
  [arXiv:1001.3103 [hep-th]].

\bibitem{Gross:1973ju}
D.~J.~Gross and F.~Wilczek,
  Phys.\ Rev.\ D {\bf 8} (1973) 3633.

\bibitem{Pickering:2001aq}
  A.~G.~M.~Pickering, J.~A.~Gracey and D.~R.~T.~Jones,
  Phys.\ Lett.\ B {\bf 510} (2001) 347 [Erratum-ibid.\ B {\bf 535}, 377 (2002)]
  [hep-ph/0104247].

\bibitem{Dudal:2012sb}
D.~Dudal and S.~P.~Sorella,
  Phys.\ Rev.\ D {\bf 86} (2012) 045005
  [arXiv:1205.3934 [hep-th]].


\bibitem{Lenz:2000zt}
F.~Lenz, J.~W.~Negele, L.~O'Raifeartaigh and M.~Thies,
  Annals Phys.\  {\bf 285} (2000) 25   [hep-th/0004200].


\bibitem{Dudal:2007cw}
D.~Dudal, S.~P.~Sorella, N.~Vandersickel and H.~Verschelde,
  Phys.\ Rev.\ D {\bf 77} (2008) 071501
  [arXiv:0711.4496 [hep-th]].

\bibitem{Dudal:2008sp}
  D.~Dudal, J.~A.~Gracey, S.~P.~Sorella, N.~Vandersickel and H.~Verschelde,
  Phys.\ Rev.\ D {\bf 78} (2008) 065047
  [arXiv:0806.4348 [hep-th]].


\bibitem{Dudal:2011gd}
D.~Dudal, S.~P.~Sorella and N.~Vandersickel,
  Phys.\ Rev.\ D {\bf 84} (2011) 065039
  [arXiv:1105.3371 [hep-th]].

\bibitem{Gracey:2010cg}
J.~A.~Gracey,
  Phys.\ Rev.\ D {\bf 82} (2010) 085032
  [arXiv:1009.3889 [hep-th]].

\bibitem{Dudal:2010tf}
D.~Dudal, O.~Oliveira and N.~Vandersickel,
  Phys.\ Rev.\ D {\bf 81} (2010) 074505
  [arXiv:1002.2374 [hep-lat]].

\bibitem{Cucchieri:2011ig}
A.~Cucchieri, D.~Dudal, T.~Mendes and N.~Vandersickel,
  Phys.\ Rev.\ D {\bf 85} (2012) 094513
  [arXiv:1111.2327 [hep-lat]].

\bibitem{Oliveira:2012eh}
O.~Oliveira and P.~J.~Silva,
  arXiv:1207.3029 [hep-lat].


\bibitem{Dudal:2012zx}
D.~Dudal, O.~Oliveira and J.~Rodriguez-Quintero,
  Phys.\ Rev.\ D {\bf 86} (2012) 105005
  [arXiv:1207.5118 [hep-ph]].


\bibitem{Fukushima:2012qa}
K.~Fukushima and K.~Kashiwa,
  arXiv:1206.0685 [hep-ph].

\bibitem{prep}
M.~Capri {\it et al.}, work in progress.


\end{thebibliography}
\end{document}